# Potentials and Challenges of Light Fidelity Based Indoor Communication System

Farooq Aftab

School of Computer and Communication Engineering

University of Science and Technology Beijing china (USTB)

farooqaftabpak@yahoo.com

## ABSTRACT

In this era of modern devices and high speed communication the issue of spectral overloading is increasing with time and becoming more serious. With the advancement in LED industry, light fidelity (Li-Fi) based indoor network is an attractive substitute for the existing radio frequency (RF) based communication networks. Because of its capability to perform dual function of lighting as well as high speed communication, Li-Fi is attracting both industrial as well as academic researchers. The Li-Fi networking paradigm offers performance enhancements which can make it an attractive backup option to be used for networking setup in Internet of things (IOT) and its indoor capabilities make it an attractive choice for large scale indoor communication in next generation wireless networking environment. This paper discusses key potentials of Li-Fi based indoor communication system and point out the challenges which indoor based Li-Fi network is facing under the knowledge of existing research work in the field of Li-Fi.

## 1 INTRODUCTION

Light fidelity (Li-Fi) is one of the new and emergent field of optical wireless communication (OWC) that provides an opportunity to move forward toward higher frequencies in EM spectrum by using a visible light spectrum. The rapid increase in the usage of LEDs and its ability to perform dual function of illumination as well as communication has provided a unique opportunity for indoor lighting and wireless communication system to go through a revolution. LEDs have an ability to switch from different intensity of light at a rapid rate, this capability can be utilize for sending information using visible light spectrum.

Apart from rapid blinking rate at different intensity of light LEDs also have high lifespan. They are energy efficient and a good option to be used for indoor because of lower generation of heat. These benefits let the LEDs to be a perfect choice for a new technology called Li-Fi which can be useful for indoor communication and can also provide a backup for wireless fidelity (Wi-Fi) technology. Li-Fi could be classified as nm-wave communication [1] because it uses higher band of frequencies in the electromagnetic spectrum for high speed data communication. Li-Fi is a complete multi user wireless network that could operate user wireless network that could operate simultaneously alongside with Wi-Fi and long term evolution (LTE) [2]. It is a green communication method because it reuses the existing lightning infrastructure.

Li-Fi can be consider as more advance networking system as compare to VLC because it can provide point to multipoint communication. This point to multipoint based communication characteristics of Li-Fi system make them different from VLC based systems because VLC only supports point to point communication system [1]. Li-Fi is a continuation of VLC technology using LEDs that can provide a proper networked based wireless system with high speed communication. It is a bi-directional multiuser communication system which can enables full user mobility because of its multiple access point formation. The fact that LEDs are natural beam formers, enables local containment of Li-Fi signals and because of the blockage of the signals by opaque walls, Co-channel interference can effectively be managed and physical layer security can be enhanced. Li-Fi can decrease the traffic bottlenecks caused due to large number of users in RF based indoor networks. Li-Fi can act as a green wireless network based technology for 5G networks [3] because it operates in the unlicensed and safe visible light frequency spectrum which can be helpful for the proper reuse of bandwidth and provides an efficient wireless solution by minimizing the capacity drainage problem of frequency spectrum.

This paper will explain in detail about Indoor implementation of Li-Fi based communication system and emphases on the research conducted by researchers on indoor based Li-Fi systems. The multiple section of this paper are organized as follows, section II consists of the elements which affect the functionality of Li-Fi based indoor network. Section III reviews the key potentials of Li-Fi based network and Section IV emphases on main challenges, weaknesses and issues of Li-Fi network at its current stage.

## 2 ELEMENTS WHICH AFFECT PERFORMANCE OF INDOOR BASED LI-FI NETWORK

Indoor based Li-Fi network has certain elements which plays an important role in the performance and efficiency of this system. In this section we will explain then one by one

### 2.1 Modulation Technique

In Li-Fi based system, multiple types of modulation schemes [4] can be useful but because of communication channel having frequency response in non-flat nature , most of the modulation techniques suffer from an undesired channel response called inter symbol interference (ISI). As most of the commonly used modulation methods such as pulse width modulation [5], pulse position modulation, ON-OFF keying, unipolar pulse amplitude modulation suffer from ISI therefore there is a need of such a scheme for Li-Fi which correlates the energy and organize itself adaptivity according to the properties of communication channel. Multicarrier modulation can provide higher data rate and it can also be useful to decrease the effect of interference and distortion but these modulation techniques are less energy efficient. OFDM [6] is most commonly used modulation technique. In OFDM signal is bipolar and its value is complex in nature. Unipolar signal can be obtained by applying a positive DC bias voltage which  can vary the amplitude of the OFDM signal. This modulation scheme is given a name DC biased optical orthogonal frequency division multiplexing (DCO-OFDM) [7]. Such schemes can be useful to implement in scenarios when a system is used to perform dual function of communication as well as illumination. This DC bias method has a drawback that it can considerably compromise efficiency of energy in whole modulation scheme. That is the reason why researchers have dedicated noteworthy efforts in designing a pure unipolar based OFDM modulation schemes. Asymmetrically clipped optical orthogonal frequency division multiplexing (ACO OFDM) can provide a solution to decrease the effects of this issue [8]. Some of the other solutions are flip OFDM and discrete multi-tone modulation (PAM-DMT) along with modulation technique based on pulse amplitude.

## 2.2 LED's Selection

Basic aim of indoor based Li-Fi wireless network is to achieve high data rates. In Li-Fi network, the selection of LEDs can plays an important role because LEDs blinking rate can affect the overall data rate of entire network. LED's parameters such as its size, ON-OFF speed and color rendering ability are also very valuable attributes. Rate of data transmission is inversely proportional to the dimension of LED's bulb. ON-OFF blinking speed of LED can also control the data rate.  Faster the rate of LED blinking higher will be the rate of data transmission. Number of LED's in a system is another factor which increases the data rate. The speed of data transmission can also be enhanced with more coverage area and better ability to accommodate more number of users. The low cost of incoherent solid state LED lighting make them suitable for deployment in indoor based Li-Fi network [9]. In Li-Fi data is encoded according to the intensity of the light emitted from light source. The data is transmitted in the sequence of 0's and 1's without modulating the amplitude and actual phase of the light wave.

LED color can also affect the data rate [10]. White color LEDs coated with phosphor can provide data rate of 1 Giga bits per second [11], similarly combination of red, green and blue color LEDs (RGB) can boost the data rate up to 3.4 Giga bits per second [12]. Incoherent LEDs of single color has reported a rate of approximately 3.5 Giga bits per second.

## 2.3 Indoor Environment model

LED's based transmitter configuration on ceiling is shown in figure 1. In section A of the figure we can see 4 different section of LED's are deployed with the ceiling of the room while Section B show us Layout model of uniform configuration. Section A layout is useful when we have small number of users using Li-Fi facility. Those users will accommodate at the place where the Li-Fi transmitter are deployed. Section B layout is useful if number of user in indoor unit are large and have to move everywhere [13].

## 2.4 LED Source Panel (LSP)

LSP is a light source which contains LED bulbs. These LED bulbs can accomplish both roles of illumination and as well as data communication. LED's bulb in a LSP can be circular or rectangular in shape. Both LED's shapes have its own advantages. Circular shaped LED's are used in Li-Fi system when we have to apply light at a fixed place while rectangular shaped chip like LED's are used when we have to disperse light at a wider area. Number of LED's in a LSP depends upon the size of LSP. If LSP has to cover Indoor environment of large specific area then it size may be large as a result it most of more number of LED's.

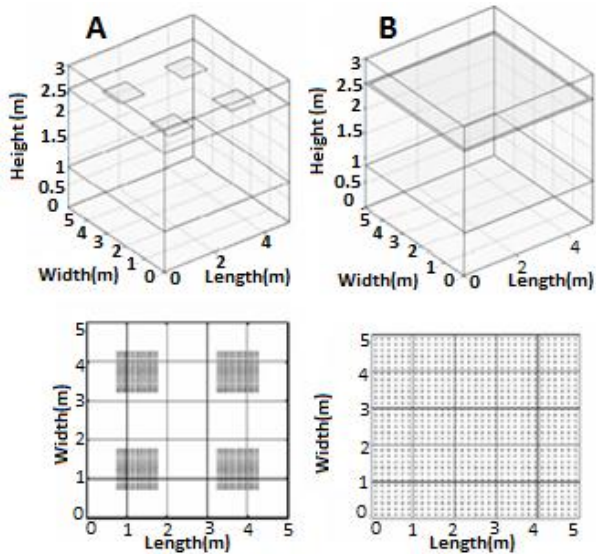

**Figure 1.** Indoor Environment Model

Total number of LSP in a room is depending upon:

1) A maximum numbers of Li-Fi user needs to accommodate in a system.

2) Total area of room.

**2.5 LSP Placement Design**

In Li-Fi, the placement of light source panel (LSP) plays an important role. LSP can set the limit of data rate because the intensity of light from LED can be controlled by using LSP. The quality of service can be managed by introducing a threshold value for indoor communication system. For every user, light intensity need to be managed in such an order that everyone can reach that threshold value for efficient communication. We propose a fixed LSP design as shown in figure 2. The covering area of LSP is adjusted in two ways [14]. Section A of figure 2 consists of fixed LSP which cover two users. In Section B we provide an approach of separate LSP for two users which is called dedicated LSP approach. In Section C we have shown dedicated LSP approach for 3 users at single place.

Fixed Single LSP approach [14] can accommodate multiple users at a time. This approach can provide a cost effective solution by providing service to multiple users in a single time slot. Dedicated LSP approach can ensure high data rate and high speed transmission of information along with a secure environment. So all these scenario based approaches can provide a tradeoff between cost and high speed.

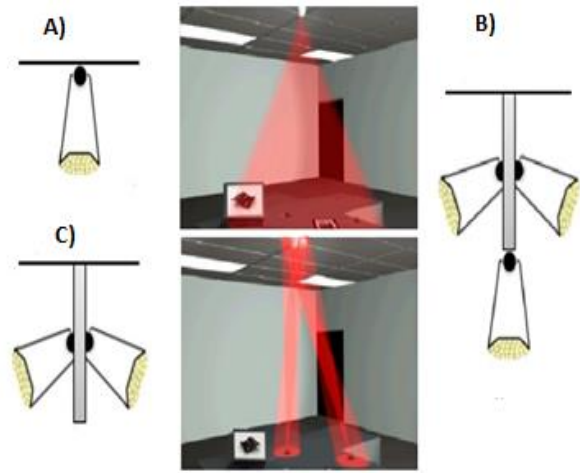

**Figure 2.** LSP in term of Covering Area

**2.6 Li-Fi channel model**

Li-Fi channel contains diffuse components as well as line of sight (LOS) components as shown in figure 3. The LOS component can be explained as [15]:

$$L = \begin{cases} \frac{(m+1)A_P}{2\pi(z^2+h^2)} g(\theta)T_s(\theta)cos^m(\phi)\cos(\theta), & \theta < \Phi_F \\ 0, & \theta \geq \Phi_F \end{cases}$$

where m is the lambertian index which can be describe as a function of radiation angle having half-intensity. $A_P$ represents the coverage area of the optical photo detector, z shows the horizontal distance between access point and optical photo detector .

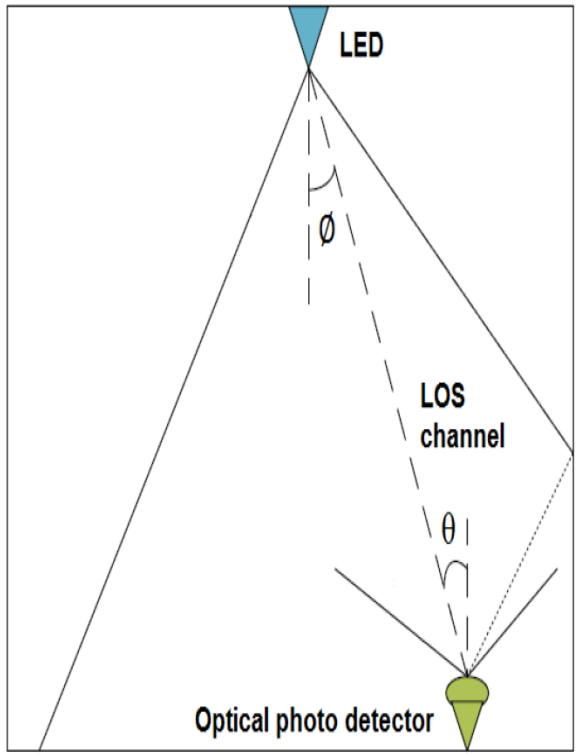

**Figure 3.** Li-Fi channel model

Here h, $\phi$, $\theta$, $\Phi$ is the height of the room, angle of irradiation, incidence and half angle of the field of view (FOV) regarding optical photo detector respectively while $T_s(\theta)$ is the gain of the optical filter and $g(\theta)$ is the concentrator gain.

## 3 KEY POTENTIALS OF LI-FI BASED NETWORK

Some of the key potentials of Li-Fi based network are given below

### 3.1 Multiple access in Li-Fi

Li-Fi can accommodate multiple users with simultaneous network access [16]. Optical space division multiple access (SDMA) can be useful which use an angle diversity transmitter. When compared with the optical time division multiple access (TDMA) technique, it has been shown that optical SDMA can achieve more throughput with in a Li-Fi network. However, such performance enhancement requires careful design of the angle diversity transmitter and time-consuming user-grouping algorithms based on exhaustive search. OFDMA provides a straight forward method for multiuser access [17] where users are served and separated by a number of orthogonal subcarriers but subcarriers with lower frequencies generally provide users with high SNR statistics. Therefore it is important in OFDMA to use appropriate user-scheduling techniques to ensure that fairness in the allocation of resources (subcarriers) is maintained. In order to enhance the throughput, non-orthogonal multiple access (NOMA) was proposed which utilize the broadcasting nature of LEDs, the performance of a Li-Fi network can be efficiently enhanced with the application of NOMA. NOMA is unique as compare to orthodox multiple access technologies because it can serve an increased number of users by using its unique resource allocation method which is non-orthogonal in nature.

## 3.2 Li-Fi based internet AP

In an indoor Li-Fi based network [18], every lighting place in a room can act as an optical access point (AP). If the distance between APs is small then the interference between them is unavoidable. This issue can considerably affect the performance of a network. To solve this issue, angle diversity based optical photo detector is proposed to mitigate the LOS interference. This optical photo detector is consists of narrow FOV along with multiple number of directional photodiode. The indoor based Li-Fi network is comprises of two parts, Li-Fi Access Point (AP) and the Li-Fi terminal.

The Local Area Network (LAN) of Li-Fi is consists of two layers, MAC layer and PHY layer. The PHY layer guarantees the efficiency of data stream whereas the MAC layer is used to manage the flow between transmitter and optical receiver. It also ensures that the data is properly transmitted and received in to the form of frames. The MAC layer uses an RJ45 port to connect with the switch. The digital module of PHY layer is used to process the transmitted data stream and LED received an encoded digital signal which is transmitted using an LED light. An optical photo detector is used to receive the beam of a light at receiver end. This optical photo detector transform optical signal back to original data signal.

## 3.3 Hybrid indoor system based on  Li-Fi and Wi-Fi

Li-Fi networks can achieve high throughput by deploying large number of APs [19].  But the spatial distribution of the data rates fluctuates due to the CCI. In order to augment the system performance and to guarantee equally high Quality of Service (QoS) among users, Wireless-Fidelity (Wi-Fi) overlay can be deployed. As Li-Fi is using a different band of frequency spectrum as compare to Wi-Fi so there is no interference among these systems. Therefore, a hybrid system consists of Wi-Fi and Li-Fi network is capable of achieving the desirable throughput. Wireless Gigabit Alliance (WiGig) is one of the latest member of Wi-Fi family can be considered for hybrid network. This latest protocol can operate on three bands of frequency and also consists of some modern advance features. By considering a hybrid network between Wi-Fi and Li-Fi, user's at all possible locations within an enlarged coverage area can benefit from significantly enhanced user throughput and QoS. This hybrid system can provide benefits of reduction of contention as a result losses of spectrum efficiency will reduce. Li-Fi system can provide offload to the present Wi-Fi system and additional benefit of coverage at dead spots can be achieve.

## 3.4 Li-Fi as Intelligent Lighting

Li-Fi system can act a smart system by giving an advance feature of power saving. The brightness level of lighting system can be controlled according to the number of users and their requirement to save power by using sensors. These sensors can be deployed to monitor multiple parameters such as intensity of light, blinking level of LED and its color. The coverage area of a LED transmitter can also be controlled by using dimming level of a LED transmitter. This intelligent lighting system can provide a smart solution to control the power consumption of LEDs. These networks can be used in smart home systems [20] where LED based lighting can provide illumination and data communication at the same time. In these smart homes devices which are used for data communication such as laptop, cellphone, and other smart devices can also perform short distance communication at high speed using visible light spectrum.

# 4 MAIN CHALLENGES FOR LI-FI BASED COMMUNICATION SYSTEM

The Li-Fi based communication system faced different kinds of challenges. These challenges limit its performance and can decrease the overall efficiency of the network. Some of the main challenges are given below:

## 4.1 LED related issues

Some of the LED related issues are

### 4.1.1   LED light ON-OFF mode

Indoor Li-Fi based communication system aims to provide illumination with communication, so ON-OFF speed of a LED plays a vital role. For a Li-Fi based system it is always compulsory to have a Light source in ON condition but it initiates main problem of how data transmission will occur when the LEDs are turned OFF. A data transmission can still be possible if brightness level of a LED transmitter is very low. The dimming level of LED bulb can be organized in such a way that a desired data rate can be achieved using light intensity. In hybrid setup, RF or infrared can be useful to provide communication in LED OFF mode but in Li-Fi based communication it is still a challenge to find a suitable solution of how communication will be possible in any undesirable situation when LEDs are in its OFF mode.

### 4.1.2   LED Junction Temperature

The management of thermal temperature is a critical design issue of high power LEDs. High junction temperature can affects spectral efficiency. Junction temperature of LED can be increase due to variation in drive current, self-heating and ambient temperature. This high junction temperature could cause degradation in power of a single with respect to time which reduces the signal to noise ratio (SNR) and degrades the lifespan of LEDs [21]. The effect could cause serious problems if array of hundreds of LEDs are connected closer to each other in a lighting system at large scale.

## 4.2 Indoor modeling issues

Some of the indoor modeling issues are

### 4.2.1   FOV Alignment

In Li-Fi network an assumption is consider before communication that transmitter and receiver have a LOS connection. The LOS connection can provide high data rates because the transmitter and receiver are aligned their FOV to maximize the channel response. Nevertheless, in real life practical scenarios, a receiver FOV can be changed and it can also move from one place to another. The change in orientation of a receiver and its mobility suggest that receiver's FOV cannot always be aligned with the transmitter. Therefore it is essential to design such techniques which can handle the scenario of FOV misalignment and provides desirable data rates. This needs modification in schemes and development of new approaches to handle this problem but designing such schemes and methods is extremely challenging and it is an important direction of future research.

### 4.2.2   Shadowing

The data rate in Li-Fi network will decline if an obstacle blocks the LOS channel as a result overall performance of the network will degrades. Not enough research is done until now to understand the indoor model and effect of shadowing on Li-Fi [22]. Shadowing could be one of the reasons of LOS channel blockage and it can produce variations in received signals therefore it is necessary to have a mechanism to provide an alternative wireless connection in a typical blockage event. It is also possible that the blockage event is of very short duration caused by the passing of obstacles or humans so it necessary to propose such a schemes and mechanisms that can provide a solution of problems such as FOV misalignment and shadowing.

### 4.2.3   Interference

In Li-Fi system light from any other energy source except of LED such as sun light or free ordinary electric light source can cause interference because it can interrupt the LOS channel between transmitter and receiver. The interruption in path of transmission will affect the data communication therefore for indoor communication new techniques are required to find solutions regarding this condition.

### 4.3 Receiver Design issues in the case of mobility

Li-Fi receivers can consist of an optical photo detector or an imaging sensor for receiving the beam of light. The photodiode is more beneficial for stationary users because in this case receiver FOV can be aligned easily to the LED. The imaging sensor has comparatively larger FOV so they can be useful for devices which support mobility but imaging sensors are less energy efficient and also produce delays in data reception as a result can decrease the overall achievable data rate. Therefore it is challenging to design such an optical receiver that can control FOV misalignment and increase robustness. Hence for both static and mobility cases, an enhancement in optical receiver design is needed to ensure high data rates along with power efficiency.

### 4.4 Li-Fi internet connectivity issues

For Li-Fi based broadband access network, it is essential that LEDs driving circuit is connected with internet [23]. The cost of internet deployment for Li-Fi and the interference of wireless connections is a limiting factor which can reduce the achievable data rate using internet. Efficient designing techniques are required to provide desirable internet connectivity speed using

LEDs at affordable deployment cost. Therefore it becomes a challenging task to propose a model which can provide internet using Li-Fi for large scale communication.

### 4.5 Up link transmission issues

A Wireless communication network is incomplete without the facility of uplink communication. In Li-Fi uplink requires that transmitter and receiver maintains a directional link during transmission. It can significantly reduce the overall throughput of the network if both devices are constantly moving. So in Li-Fi it is also a challenge that how the uplink traffic in a network will be operate. The radio frequency and infrared can be considered for transmitting uplink data in Li-Fi network but still more innovative ideas are require for solving the uplink issues in Li-Fi networks.

### 4.6 Connectivity and Coverage area issues

It is necessary for a Li-Fi system to maintain continuous and high speed connectivity within a coverage area of a Li-Fi cell and between the Li-Fi cells. So advance schemes for link layer are require which can maintain rate adaptation and frame aggregation to cope up with connectivity issues. In a Li-Fi based network it is essential that smooth handover of devices as well as handover of technology will occur for efficient communication in advanced Li-Fi based system.

### 4.7 Security threats

In recent research [24] it is proposed by the researchers that Li-Fi network can also suffer from security threats. An attacker may be present inside or outside a room can perform eavesdropping using the light signals. These signals can be obtained from gap between floor and door, cracks inside flooring or from partially shielded windows. This threat indicates that more research is required to understand and resolve the security issues and privacy concerns of Li-Fi network.

## 5 CONCLUSION

The vision behind Li-Fi technology is to provide a high speed data communication using visible light spectrum and its future looks bright for indoor implementation because of rapid increase of LEDs for indoor lighting. With LEDs expected to slowly replace the traditional lighting system, Li-Fi is foreseen to be gradually implemented into general lighting infrastructures which will give rise to several beneficial applications. Broadband internet can also be accessible using same lighting system which provides us illumination in our daily life. Li-Fi has a potential of large scale implementation and this technology can be improve with time which attracts many companies , designers and researchers to keep working for the practical implementation of Li-Fi network for indoor communication. Indeed, research teams are working on multiple schemes, algorithms, indoor models and new techniques to compensate for the limitation of Li-Fi network. System load balancing can be achieved from Hybrid Li-Fi and Wi-Fi based networks. The limitation of Li-Fi system such as sensitivity to the line of sight connection and non-uniform spatial distribution of data rates due to co-channel interference (CCI) need to be controlled in order to attain high speed communication with desirable data rate. As a result of this technology every LED bulb can act as a hotspot to transmit wireless data and the world will move toward the cleaner, greener, safer and brighter future.

**Author**

**Farooq Aftab** received his BS Telecommunication Engineering degree in 2013 from Foundation University, Islamabad, Pakistan. Currently he is pursuing master degree in Information and Communication Engineering from University of Science and Technology Beijing, China. His research area is Mobile ad-hoc network (Manets) and Light fidelity (Li-Fi).

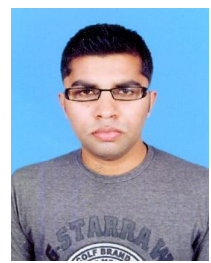